%% file: LIU_WCL_2025_2824.tex
\documentclass[journal]{IEEEtran}
\usepackage{algorithm,algorithmic}
\usepackage{amsmath,amssymb,mathrsfs}
\usepackage{array}
\usepackage{balance}
\usepackage{booktabs}
\usepackage{cite}
\usepackage{epsfig}
\usepackage{easyReview}
\usepackage{fancyhdr}
\usepackage{float}
\usepackage[nomain,acronym,automake,toc,nopostdot]{glossaries}
\usepackage{graphicx,color}
\usepackage{hyperref}
\usepackage{mathtools}
\usepackage{multirow}
\usepackage{psfrag}
\usepackage{stfloats}
\usepackage{subfigure}

\newtheorem{cor}{Corollary}

\newtheorem{rem}{Remark}

\newtheorem{thm}{Theorem}

\makeglossaries
\input{Acronyms}

\definecolor{sblue}{RGB}{0,51,120}
\definecolor{sred}{RGB}{200,51,130}

\newcommand{\algref}[1]{\textbf {Algorithm \ref{#1}}}
\newcommand{\figref}[1]{Fig. \ref{#1}}

\newcommand{\thmref}[1]{{\it Theorem \ref{#1}}}
\newcommand{\corref}[1]{{\it Corollary \ref{#1}}}

\newcommand{\appref}[1]{{\textsc{Appendix} \ref{#1}}}
\renewcommand{\eqref}[1]{(\ref{#1})}

\ifCLASSINFOpdf
\else
\fi

\begin{document}
\title{Phase-Rotated Symbol Spreading for Scalable \\ Rydberg Atomic-MIMO Detection}
\author{Jiuyu Liu, Yi Ma, and Rahim Tafazolli
\thanks{Jiuyu Liu, Yi Ma, and Rahim Tafazolli are with the 5/6GIC, Institute for Communication Systems, University of Surrey, Guildford, United Kingdom, GU2 7XH, e-mails: (jiuyu.liu, y.ma, r.tafazolli)@surrey.ac.uk.}
}
\maketitle

\begin{abstract}
	Multiple-input multiple-output (MIMO) systems using Rydberg atomic (RA) receivers face significant scalability challenges in signal detection due to their nonlinear signal models.
	This letter proposes phase-rotated symbol spreading (PRSS), which transmits each symbol across two consecutive time slots with an optimal $\boldsymbol{\pi/2}$ phase offset.
	PRSS enables reconstruction of an effective linear signal model while maintaining spectral efficiency and facilitating the use of conventional RF-MIMO detection algorithms.
	Simulation results demonstrate that PRSS achieves greater than 2.5 dB and 10 dB bit error rate improvements compared to current single-transmission methods when employing optimal exhaustive search and low-complexity sub-optimal detection methods, respectively.
\end{abstract}

\begin{IEEEkeywords}
Rydberg atomic (RA) receiver, multiple-input multiple-output (MIMO), scalable RA-MIMO detection.
\vspace{-0.9em}
\end{IEEEkeywords}

\section{Introduction}\label{sec1}
Rydberg atomic (RA) receivers use highly excited alkali atoms (e.g., rubidium, cesium) for radio wave detection, which fundamentally differs from \gls{rf} antennas \cite{Zhang2023d}.
They offer several advantages: self-calibration based on physical constants, continuous DC-to-THz frequency tunability, and theoretically quantum-limited sensitivity exceeding $20$ dB compared to RF antennas \cite{Bussey2022, Fancher2021}.
RA receivers produce Rabi frequency readouts proportional to the square of the RF signal envelope, introducing phase ambiguity that initially limited implementations to low-rate amplitude modulation \cite{Liu2025b}.
Since 2019, heterodyne principles have enabled phase-modulated symbol transmission by injecting reference RF signals \cite{Simons2019}.
Current \gls{siso} prototypes demonstrate Mbps-level data rates, supporting phase modulation schemes like \gls{qam} \cite{Anderson2021}.

Recent works have extended from \gls{siso} to \gls{mimo} configurations for higher spectral efficiency \cite{Cui2024a, Liu2025a, Cui2025}. 
However, RA-MIMO detection faces significant scalability challenges due to the inherent nonlinear signal model \cite{Cui2025, Atapattu2025, Xiao2025}.
This fundamentally alters RA-MIMO signal detection, making traditional linear MIMO algorithms inapplicable.
Current solutions include optimal \gls{ml} detectors with computational complexity that grows exponentially with the number of users \cite{Cui2025}.
The expectation maximization Gerchberg-Saxton (EM-GS) algorithm offers an iterative alternative but requires the number of RA receivers to exceed the number of users tenfold for near-optimal performance \cite{Cui2025}.
Existing methods accept the nonlinear RA-MIMO signal model as given and attempt to develop sophisticated receiver-side detection algorithms \cite{Dong2023}.
However, in wireless communication systems, transmitter-side modifications can be leveraged to address this challenge.

This motivates us to propose \gls{prss}, a transmitter-receiver co-design approach that directly tackles the fundamental nonlinearity challenge in RA-MIMO systems.
PRSS transmits each information symbol with two distinct phase rotations across consecutive time slots, enabling each RA receiver to independently recover effective received signals containing both positive and negative values, thereby directly solving the nonlinearity problem inherent in Rabi frequency readouts.
Through theoretical analysis, we derive that the optimal phase offset is $\pi/2$, which achieves maximum \gls{snr}.
By applying this technique across all RA receivers, we reconstruct a linear MIMO signal model equivalent to that of RF-MIMO systems, enabling direct use of established RF-MIMO detection algorithms.

While PRSS requires two time slots per symbol, it maintains comparable spectral efficiency to single-transmission RA-MIMO systems since the inherent nonlinearity in RA receivers already reduces spectral efficiency by approximately half relative to linear RF systems.
In our simulations, we normalize data rates by employing $16$-QAM modulation for PRSS versus $4$-QAM for single-transmission approaches, providing a fair comparison since PRSS transmits over two time slots.
The performance advantages are significant: PRSS with optimal detection provides \gls{ber} gains exceeding $2.5$ dB compared to existing single-transmission methods.
Furthermore, integration with computationally efficient sub-optimal detectors like \gls{zf} achieves performance gains greater than $10$ dB relative to current EM-GS method.

\section{RA-MIMO Signal Model and Problem Statement} \label{sec2}
This section introduces the principles of RA receivers and their inherent nonlinear signal model, then presents the motivation and design objectives for the proposed PRSS method.

\subsection{RA-MIMO Signal Model}
We begin with an example of a system with a single RA receiver, where alkali atoms such as rubidium and cesium are contained within vapor cells \cite{Simons2019}.
Laser systems are employed to excite these atoms to high-energy Rydberg states.
When these lasers satisfy the \gls{eit} condition, the probe laser experiences minimal absorption, resulting in a narrow transparency window with a single peak in the optical transmission spectrum.
When a specific RF electric field is applied, the resulting transmission spectrum splits into two peaks, a phenomenon known as Autler–Townes (AT) splitting.
The frequency separation between these two peaks is called the Rabi frequency, which is directly proportional to the RF field amplitude, as follows \cite{Zhang2023d}
\begin{equation}
	\Omega = \bigg|\dfrac{\mu E_{\textsc{rf}}}{\hbar}\bigg|,
\end{equation}
where $E_{\textsc{rf}}$ denotes the RF electric field amplitude, $\mu$ represents the dipole moment for the Rydberg state transition, and $\hbar$ is the reduced Planck constant.
The atomic response time is much slower than the carrier frequency, effectively acting as a natural low-pass filter to directly read out the baseband signal.
Let $x_{n}(t)$ denote the complex slow-varying baseband signal transmitted by the $n$-th RF antenna, the final readout at an RA receiver can be described as follows \cite{Zhang2023d}
\begin{equation} \label{eqn05110921}
	z(t) = \Bigg|\sum_{n=1}^{N}h_{n} x_{n}(t) + v(t)\Bigg|,
\end{equation}
where $h_n\in \mathbb{C}$ denotes the complex channel gain from the $n$-th antenna, and $v(t)$ represents the noise combination of blackbody radiation noise, photon shot noise, and electronic thermal noise.
The amplitude-only observation initially prevented the use of high-rate phase-modulated symbols such as \gls{qam}.

To overcome this limitation, researchers in 2019 proposed deploying a \gls{lo} to generate a reference RF signal that is injected along with the information RF signal \cite{Simons2019}.
Building upon this heterodyne principle, recent works have extended RA receivers from \gls{siso} to \gls{mimo} configurations to achieve higher spectral efficiency.
For an RA-MIMO system with $M$ receivers and $N$ user antennas, this leads to the following uplink signal model \cite{Cui2025, Atapattu2025, Xiao2025}
%\footnote{Unlike some RA-MIMO signal models that leverage approximate linearization, i.e., $\mathbf{z} \approx \boldsymbol{\Theta}\mathbf{H}\mathbf{x}+\mathbf{v}$ with $\boldsymbol{\Theta}$ containing quantum transduction gains and phases, we focus on this fundamental signal model direct developed form direct AT-splitting measurements.}
\begin{equation} \label{eqn03160331}
	\mathbf{z}^{(1)} = \big|\mathbf{H}\mathbf{x} + \mathbf{v}^{(1)} + \mathbf{r}\big|,
\end{equation}
where the operator $|\cdot|$ represents the element-wise amplitude operation,
$\mathbf{z}^{(1)} \in \mathbb{R}^{M \times 1}$ denotes the amplitude observations; 
$\mathbf{H} \in \mathbb{C}^{M \times N}$ represents the random channel matrix,
$\mathbf{x} \in \mathbb{C}^{N \times 1}$ contains the transmitted signals from users, with each element drawn from a finite alphabet set $\mathbb{X}$ with equal probability, $\mathbf{r} \in \mathbb{C}^{M\times 1}$ is the LO-induced reference signal, and $\mathbf{v}^{(1)} \sim \mathcal{CN}(0, \sigma_{v}^{2}\mathbf{I})$ represents \gls{awgn} with $\mathbf{I}$ denoting the identity matrix.

\subsection{Motivation and Design Objectives} \label{sec02b}
The nonlinear signal model of RA-MIMO systems poses significant scalability challenges for large-scale detection.
PRSS addresses this limitation through a simple yet effective approach: transmitting each information symbol twice with different phase rotations.
Specifically, after transmitting symbol $\mathbf{x}$ in the first time slot (resulting in observation $\mathbf{z}^{(1)}$), each user retransmits the same symbol with a phase offset $\phi$ in the second time slot, as follows
\begin{equation} \label{eqn09080403}
	\mathbf{z}^{(2)} = \big|\mathbf{H}\mathbf{x}e^{j\phi} + \mathbf{v}^{(2)} + \mathbf{r}\big|,
\end{equation}
where the channel matrix $\mathbf{H}$ and reference signal $\mathbf{r}$ remain constant across both time slots.
The key insight is that these two amplitude observations, combined with the known reference signal, provide sufficient information to reconstruct the complex-valued received signal $\mathbf{s} = \mathbf{Hx}$.
Once $\mathbf{s}$ is estimated, the RA-MIMO system effectively operates with a linear signal model, enabling the direct application of conventional RF-MIMO detection algorithms.

This work addresses three main objectives: deriving the optimal phase offset $\phi$ that minimizes the estimation error of $\mathbf{s}$, analyzing the spectral efficiency of PRSS-enabled RA-MIMO systems, and evaluating the computational complexity compared to existing nonlinear detection methods.

\section{PRSS-Assisted RA-MIMO Detection} \label{sec03}
This section develops the PRSS method in three steps: optimizing the phase offset for maximum estimation accuracy, constructing an effective linear signal model, and analyzing performance and computational complexity.

\subsection{Optimization and Implementation of PRSS}\label{sec03a}
The core challenge is estimating the complex signal $\mathbf{s} = \mathbf{Hx}$ from two amplitude-only observations.
Since each RA receiver operates independently, we can analyze this problem element-wise. For the $m$-th receiver, the two observations are
\begin{subequations} \label{eqn08350403}
	\begin{align}
		z_{m}^{(1)} &= \big|s_{m} + v_{m}^{(1)} + r_{m}\big|; \\
		z_{m}^{(2)} &= \big|s_{m} e^{j\phi} + v_{m}^{(2)} + r_{m}\big|,
	\end{align}
\end{subequations}
where $s_m$ is the desired signal component, $v_m^{(1,2)}$ are noise terms, and $r_m$ is the reference signal from the local oscillator.

\textbf{Key Assumption:} The reference signal is much stronger than the information signal, i.e., $|r_{m}| \gg |s_{m} + v_{m}^{(1,2)}|$\footnote{\label{foot1}This assumption is readily achievable: the \gls{lo} is placed within $1$ meter of RA receivers while users usually transmit from beyond $10$ meters, providing over $35$ dB pathloss difference in typical sub-6 GHz scenarios \cite{3GPP2022}.
When combined with high-gain horn antennas for the \gls{lo}, the total power difference can exceed $45$ dB \cite{Balanis2016}.}.

Under this assumption, we can linearize the amplitude measurements using Taylor expansion:
\begin{subequations} \label{eqn10460304}
	\begin{align}
		z_{m}^{(1)}\ &\approx |r_{m}| + \Re \big\{u_{m} (s_{m} + v_{m}^{(1)}) \big\}; \\
		z_{m}^{(2)}\ &\approx |r_{m}| + \Re \big\{u_{m} (s_{m} e^{j\phi} + v_{m}^{(2)}) \big\},
	\end{align}
\end{subequations}
where $u_{m} = {r_{m}^{*}}/{|r_{m}|}$ normalizes the reference signal phase.

Effective Observations: By subtracting the known reference amplitude $|r_m|$, we obtain
\begin{subequations} \label{eqn10190401}
	\begin{align}
		y_{m}^{(1)} &= z_{m}^{(1)} - |r_{m}| \approx \Re\big\{u_{m}s_{m}\big\} + \Re\{u_{m}v_{m}^{(1)}\}; \\
		y_{m}^{(2)} &= z_{m}^{(2)} - |r_{m}| \approx \Re\big\{u_{m}s_{m}e^{j\phi}\big\} + \Re\{u_{m}v_{m}^{(2)}\}.
	\end{align}
\end{subequations}

These equations show that $y_m^{(1)}$ and $y_m^{(2)}$ provide two different linear projections of the complex signal $s_m$. The phase offset $\phi$ controls the second projection direction, directly affecting how accurately we can reconstruct $s_m$ from these two real-valued measurements.
Since estimation accuracy impacts the final detection performance, we need to find the optimal phase offset $\phi$ that minimizes the estimation error of $\mathbf{s}$, as derived in the following theorem.

\begin{thm} \label{thm01}
	Given the effective observations $y_{m}^{(1)}$ and $y_{m}^{(2)}$, the optimal phase offset $\phi \in (-\pi, \pi]$ that minimizes the least-squares estimation error of $s_m$ is as follows
	\begin{equation} \label{eqn09010331}
		\phi^{\star} = \underset{\phi}{\arg \min} |\hat{s}_{m} - s_{m}|^{2} = \pm \dfrac{\pi}{2}, \quad \forall m.
	\end{equation}
	For $\phi^{\star} = \frac{\pi}{2}$, the optimal estimate of $s_{m}$ is as follows
	\begin{equation}\label{eqn12060304a}
		\hat{s}_{m} = u_{m}^{*}\big[y_{m}^{(1)}-jy_{m}^{(2)}\big]. 
	\end{equation}
	The case $\phi^{\star} = -\frac{\pi}{2}$ yields equivalent performance and is omitted for brevity.
\end{thm}

\begin{IEEEproof}
	See \appref{appthm01}.
\end{IEEEproof}

\thmref{thm01} demonstrates that the optimal phase offset is $\pi/2$, which provides maximum information for reconstructing the complex signal $s_m$ from two real-valued observations.
According to \eqref{eqn10190401}, the values $y_{m}^{(1)}$ and $y_{m}^{(2)}$ in \eqref{eqn12060304a} can be approximated from the amplitude measurements, leading to the following practical implementation:

\begin{cor} \label{cor01}
	Given $\mathbf{z}^{(1)}$, $\mathbf{z}^{(2)}$ and reference signal $\mathbf{r}$, the estimate of $\mathbf{s}$ for optimal phase offset $\phi^{\star} = \frac{\pi}{2}$ is:
	\begin{IEEEeqnarray}{ll}
		\widehat{\mathbf{s}}\ &= \mathbf{u}^{H} \odot\big[\big(\mathbf{z}^{(1)}-|\mathbf{r}|\big)-j\big(\mathbf{z}^{(2)}-|\mathbf{r}|\big)\big], \label{eqn10450401} \\ 
		&\approx \mathbf{H}\mathbf{x} + \mathbf{v}_{\text{e}}, \label{eqn10340327}
	\end{IEEEeqnarray}
	where $\odot$ denotes element-wise multiplication, and $\mathbf{v}_{\text{e}} \sim \mathcal{CN}(0, \sigma_{v_{\text{e}}}^{2} \mathbf{I})$ is the effective noise with the same variance as the original noise $\mathbf{v}$, i.e., $\sigma_{v_{\text{e}}}^{2} = \sigma_{v}^{2}$.
\end{cor}

\begin{IEEEproof}
	Substituting $u_{m} = {r_{m}^{*}}/{|r_{m}|}$ and the approximations from \eqref{eqn10190401} into \eqref{eqn12060304a}, then expressing in vector form yields \eqref{eqn10450401}. The case $\phi^{\star} = -\frac{\pi}{2}$ follows analogously (see \appref{appthm01}).
	The unit normalization and orthogonal transformation (for $\phi^{\star} = \pm \frac{\pi}{2}$) preserve the noise variance, yielding $\sigma_{v_{\text{e}}}^{2} = \sigma_{v}^{2}$.
\end{IEEEproof}

\begin{rem}[Interaction with Channel Estimation] \label{rem01}
PRSS reconstructs the received signal $\mathbf{s} = \mathbf{Hx}$ jointly  	without requiring separate knowledge of $\mathbf{H}$.
	The channel matrix $\mathbf{H}$ is only required during the symbol detection stage, where the detection procedure is identical to that of conventional RF-MIMO systems.
	As shown in (11), the reconstructed signal model is equivalent to the RF-MIMO signal model.
	Therefore, although channel estimation errors degrade detection performance, this impact affects PRSS-assisted RA-MIMO and RF-MIMO systems equivalently.
\end{rem}

\begin{rem}[Spectral Efficiency Analysis]\label{rem02}
	Let $C_{\textsc{rf}}$ denote the channel capacity of RF-MIMO systems with linear signal model $\mathbf{y} = \mathbf{H}\mathbf{x}+\mathbf{v}$.
	For PRSS-assisted RA-MIMO, the dual-slot transmission mechanism results in spectral efficiency $C_{\textsc{prss}} \approx \frac{1}{2}C_{\textsc{rf}}$.
	In contrast, as derived in \cite{Cui2024a}, single-shot RA-MIMO systems achieve spectral efficiency $C_{\textsc{ra}} \approx \frac{1}{2}C_{\textsc{rf}}$ due to real-valued observations that inherently halve the information capacity.
	Although these two approaches employ fundamentally different mechanisms (dual-slot transmission versus real-valued detection), both incur equivalent spectral efficiency loss relative to RF-MIMO systems.
	Consequently, PRSS achieves comparable spectral efficiency to single-shot RA-MIMO, yielding $C_{\textsc{prss}} \approx C_{\textsc{ra}} \approx \frac{1}{2}C_{\textsc{rf}}$.
\end{rem}

\begin{rem}[Optimality for Non-Gaussian Noise] \label{rem03}
	The optimality of $\varphi^{\star} = \pm\frac{\pi}{2}$ extends to any noise with finite variance and zero mean under the least-squares criterion, since \eqref{eqn08390619} depends only on noise variance $\sigma^2_v$ rather than the specific distribution.
	In practice, the noise may exhibit other distributions.
	For instance, photon shot noise follows a Poisson distribution, which can no longer be approximated as Gaussian under low laser intensity conditions \cite{Cui2025}.
	The optimality of $\phi$ in such cases represents an interesting direction for future research.
\end{rem}

The equivalent linear model \eqref{eqn10340327} enables direct application of established RF-MIMO detection algorithms. For instance, the optimal \gls{ml} detection approach is formulated as \cite{Liu2023}
\begin{equation} \label{eqn03130401}
	\mathbf{x}_{\textsc{ml}} = \underset{\mathbf{x}\in \mathbb{X}^{N}}{\arg \min} \|\widehat{\mathbf{s}} - \mathbf{H}\mathbf{x}\|^{2},
\end{equation}
where $\|\cdot\|$ denotes the Euclidean norm.
ML detection performs an exhaustive search over the constellation set to find the symbol vector closest to $\widehat{\mathbf{s}}$.
However, its exponentially growing complexity becomes prohibitive for large MIMO systems.
Alternatively, \gls{zf} detector offers reduced complexity \cite{Liu2024b}:
\begin{equation} \label{eqn03140401}
	\mathbf{x}_{\textsc{zf}} = \mathcal{Q}_{\mathbb{X}}\big([\mathbf{H}^{H}\mathbf{H}]^{-1}\mathbf{H}^{H}\widehat{\mathbf{s}}\big),
\end{equation}
where $[\mathbf{H}^{H}\mathbf{H}]^{-1}\mathbf{H}^{H}$ is the channel equalizer, and $\mathcal{Q}_{\mathbb{X}}(\cdot)$ maps each equalized symbol to the nearest constellation point.

\begin{algorithm}[t]
	\small \caption{PRSS-assisted RA-MIMO detection ($\phi = \frac{\pi}{2}$)} 
	\begin{algorithmic}[1]\label{alg02}
		\REQUIRE 
		$\mathbf{z}^{(1)}$, $\mathbf{z}^{(2)}$: amplitude observations over two time slots for transmitted symbols $\mathbf{x}$ and $j\mathbf{x}$, respectively;\\
		$\mathbf{r}$: LO-induced reference signal vector.
		\ENSURE 
		Estimated transmitted symbols $\widehat{\mathbf{x}}$.
		\STATE Compute effective received signal $\widehat{\mathbf{s}}$ according to \eqref{eqn10450401};
		\STATE Apply conventional MIMO detection:
		\IF{ML detection}
		\STATE $\widehat{\mathbf{x}} \leftarrow \arg\min_{\mathbf{x}\in \mathbb{X}^{N}} \|\widehat{\mathbf{s}} - \mathbf{H}\mathbf{x}\|^2$ (see \eqref{eqn03130401});
		\ELSIF{ZF detection}
		\STATE $\widehat{\mathbf{x}} \leftarrow \mathcal{Q}_{\mathbb{X}}\big([\mathbf{H}^{H}\mathbf{H}]^{-1}\mathbf{H}^{H}\widehat{\mathbf{s}}\big)$ (see \eqref{eqn03140401});
		\ENDIF
		\RETURN $\widehat{\mathbf{x}}$.
	\end{algorithmic}
\end{algorithm}

\algref{alg02} demonstrates PRSS implementation with optimal phase rotation $\phi^{\star} = \frac{\pi}{2}$, transmitting $j\mathbf{x}$ in the second time slot.
ML and ZF detection exemplify how PRSS enables direct application of existing RF-MIMO algorithms, with compatibility extending to other advanced detectors \cite{Albreem2019}.

\textbf{Complexity Analysis:} \label{sec03b}
The computational complexity of PRSS detection consists of two components: effective signal model reconstruction and symbol detection.
The reconstruction step using \corref{cor01} has linear complexity $\mathcal{O}(M)$.
For symbol detection: \gls{zf} requires $\mathcal{O}(N^{3})$ operations, while \gls{ml} needs $\mathcal{O}(\mathcal{J}^{N})$ operations, where $\mathcal{J}$ denotes the constellation size.
In comparison, existing single-transmission RA-MIMO approaches require either $\mathcal{O}(\mathcal{J}^{N})$ complexity for optimal exhaustive search or $\mathcal{O}(KN^3)$ for the EM-GS algorithm, with $K$ the iteration number \cite{Cui2025}.
Therefore, although the reconstruction step exhibits $\mathcal{O}(M)$ complexity and requires additional hardware resources, this requirement is negligible relative to the complexity inherent in symbol detection step.

Both PRSS and existing RA-MIMO detection methods rely on the heterodyne method for signal detection. 
The fundamental distinction is that existing solutions such as EM-GS directly solve the nonlinear problem from single-slot observations, whereas PRSS transforms it into an equivalent linear problem through transmitter-side dual-slot phase rotation.

\section{Numerical and Simulation Results} \label{sec04}
The objectives of this section are to verify the optimality of the rotational angle $\phi$ derived in \thmref{thm01} and to demonstrate the performance advantages of the proposed PRSS approach.
The following four benchmarks are considered:

ML (RF receiver): The optimal detection method for MIMO systems using RF receivers with perfect phase information.

ZF (RF receiver): The sub-optimal ZF detection for MIMO systems using RF receivers with perfect phase information.

ML (Single $\mathbf{z}$): The optimal exhaustive search based method for RA-MIMO receivers with a single transmission \cite{Cui2025}.

EM-GS (Single $\mathbf{z}$)\footnote{EM-GS requires $M \gg N$ for reliable convergence, as demonstrated in \cite{Cui2025} with highly asymmetric configurations such as $36 \times 3$ and $100 \times 6$ MIMO systems. This constraint limits its scalability for practical deployment.}: The sub-optimal method for current RA-MIMO systems \cite{Cui2025}.
The iteration number is set to be $50$.

\textit{System Setup:}
The \gls{rsr} is defined as $\text{RSR} \triangleq \frac{\mathbb{E}\{|r_{m}|^{2}\}}{\mathbb{E}\{|h_{m,n}x_{n}|^{2}\}}$ \cite[(37)]{Cui2025}.
The channel matrix $\mathbf{H}$ is assumed to be known at the receiver, with each element following Rayleigh fading, i.e., $h_{m,n} \sim \mathcal{CN}(0,1/N)$.
For RA receiver parameters, we use the same configuration as \cite{Cui2025}: rubidium (Rb) atoms with Rydberg levels $52D_{5/2}$ and $53P_{3/2}$ detecting $5$ GHz RF signals.
The carrier frequency selection directly impacts achievable RSR through frequency-dependent pathloss exponents \cite{3GPP2022}.
Random atomic dipole orientations introduce channel phase variations, with polarization directions randomly sampled from unit circles perpendicular to incident angles
Despite these phase variations, the effective channel statistics preserve the Rayleigh fading distribution.
Finally, quantum shot noise under strong laser illumination is approximated as Gaussian-distributed by invoking the central limit theorem.
For fair comparison, data rates are normalized across all approaches.
RF-MIMO and single-transmission RA-MIMO systems use $4$-QAM modulation, while PRSS employs $16$-QAM.
Detection performance is evaluated using \gls{ber} with equal total energy per bit, where PRSS energy per bit is set to half that of single-transmission approaches.

\textit{Case Study 1:}
\begin{figure}[t]
	\centering
	\includegraphics[width=5.2cm]{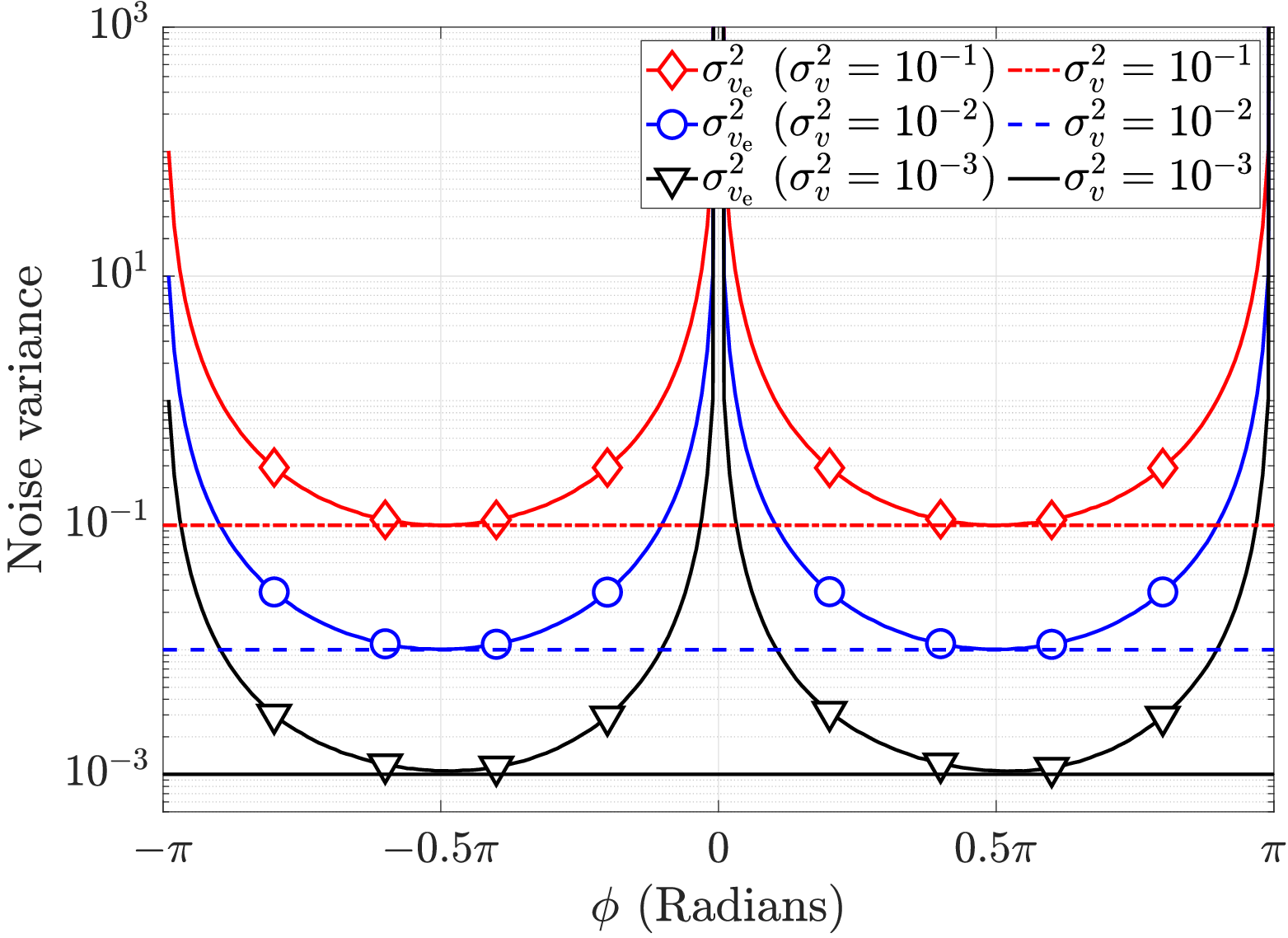}
	\caption{\label{fig02} The relationship between $\sigma_{v_{\text{e}}}^{2}$ and $\phi$ is illustrated for RSR $= 30$ dB. It can be observed that $\sigma_{v_{\text{e}}}^{2}$ achieves its minimum value at $\phi^{\star} = \pm \frac{\pi}{2}$, which is equivalent to the original noise variance $\sigma_{v}^{2}$.}
	\vspace{-1em}
\end{figure}
\begin{figure}[t]
	\centering
	\includegraphics[width=5.2cm]{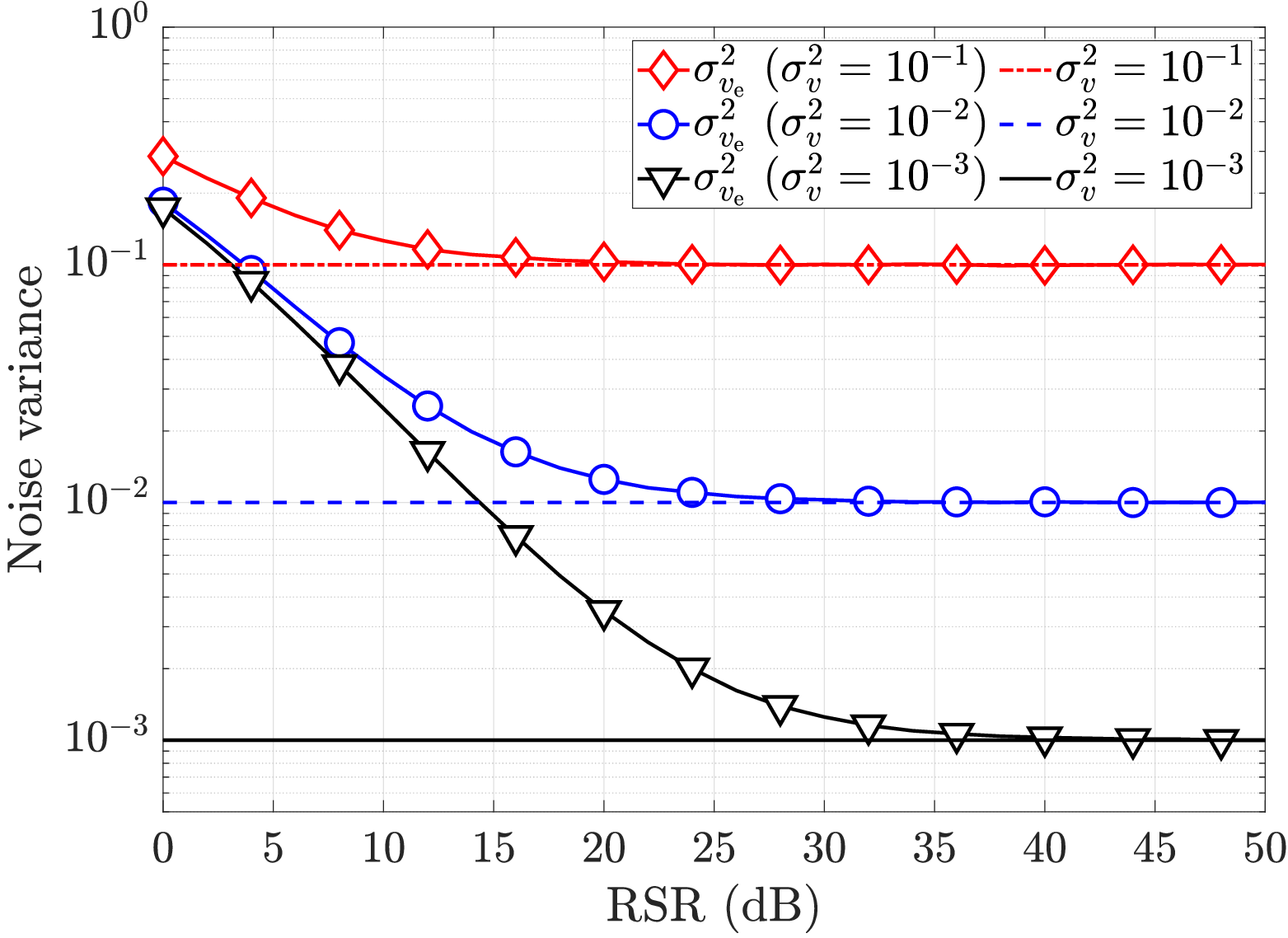}
	\caption{\label{fig022} Effective noise variance versus RSR. At low RSR levels, the Taylor expansion introduces noise amplification, particularly for smaller $\sigma_{v}^{2}$. At high RSR levels, no observable noise amplification occurs.}
	\vspace{-1em}
\end{figure}
In this case study, the following aspects are demonstrated: \textit{1)} the optimality of $\phi^{\star}=\pm\frac{\pi}{2}$, and \textit{2)} the necessity of a strong LO-induced reference signal.
The variance of the effective noise $\mathbf{v}_{\text{e}}$ is defined as:
\begin{equation}
	\sigma_{v_{\text{e}}}^{2} = \dfrac{\mathbb{E}\{\|\widehat{\mathbf{s}}-\mathbf{s}\|^{2}\}}{M},
\end{equation}
\figref{fig02} illustrates the relationship between $\sigma_{v_{\text{e}}}^{2}$ and $\phi$ for RSR $= 30$ dB.
It can be observed that the minimum effective noise variance is achieved at $\phi^{\star}=\pm \frac{\pi}{2}$, where $\sigma_{v_{\text{e}}}^{2}$ equals the original noise variance $\sigma_{v}^{2}$.
Furthermore, investigating the impact of RSR is of significant importance.
As illustrated in \figref{fig022}, the first-order Taylor expansion introduces considerable noise amplification under low RSR conditions, with this effect being more pronounced for scenarios characterized by smaller $\sigma_{v}^{2}$.
As RSR increases, $\sigma_{v_{\text{e}}}^{2}$ converges to $\sigma_{v}^{2}$, thereby providing empirical validation of the theoretical analysis.
It should be noted that at RSR $= 45$ dB, no observable noise amplification occurs across all noise levels, which represents an achievable operating condition, as discussed in footnote \ref{foot1}.

\begin{figure}[t]
	\centering
	\includegraphics[width=5.2cm]{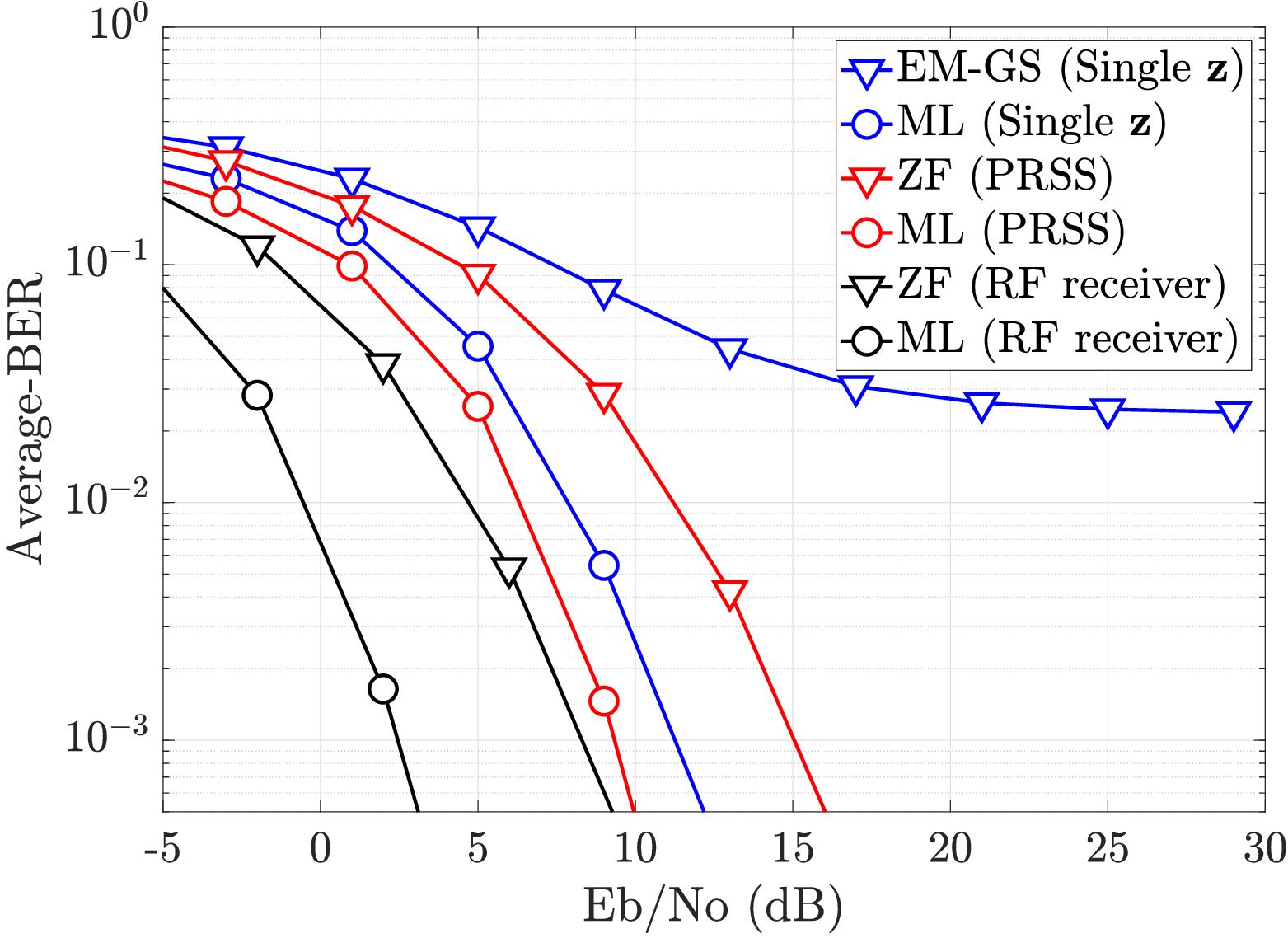}
	\caption{\label{fig03} Comparison between PRSS and four benchmarks in $8 \times 4$ MIMO systems. Benchmarks use $4$-QAM while PRSS uses $16$-QAM. ML (PRSS) achieves approximately $3$ dB gain over ML (single $\mathbf{z}$) at a \gls{ber} of $10^{-3}$.}
	\vspace{-1em}
\end{figure}

\textit{Case Study 2:}
This study evaluates the detection performance of the proposed PRSS against four benchmarks in small-scale ($M = 8$; $N = 4$) MIMO systems.
The \gls{rsr} is fixed at $26$ dB for all results in \figref{fig03}.
The results demonstrate that ML detection with PRSS achieves approximately $3$ dB performance advantage over ML detection with single transmission at a \gls{ber} of $10^{-3}$.
This indicates that the nonlinearity in RA receivers causes greater performance degradation than the overhead of dual-slot transmission.
PRSS shows approximately $7$ dB performance disadvantage compared to RF-MIMO systems.
However, RA receivers can provide more than $20$ dB sensitivity gain compared to RF receivers \cite{Bussey2022}.
Moreover, EM-GS achieves near-optimal performance only in highly asymmetric MIMO configurations \cite{Cui2025}, with performance degrading significantly as system load increases.

\textit{Case Study 3: }
\begin{figure}[t]
	\centering
	\includegraphics[width=5.2cm]{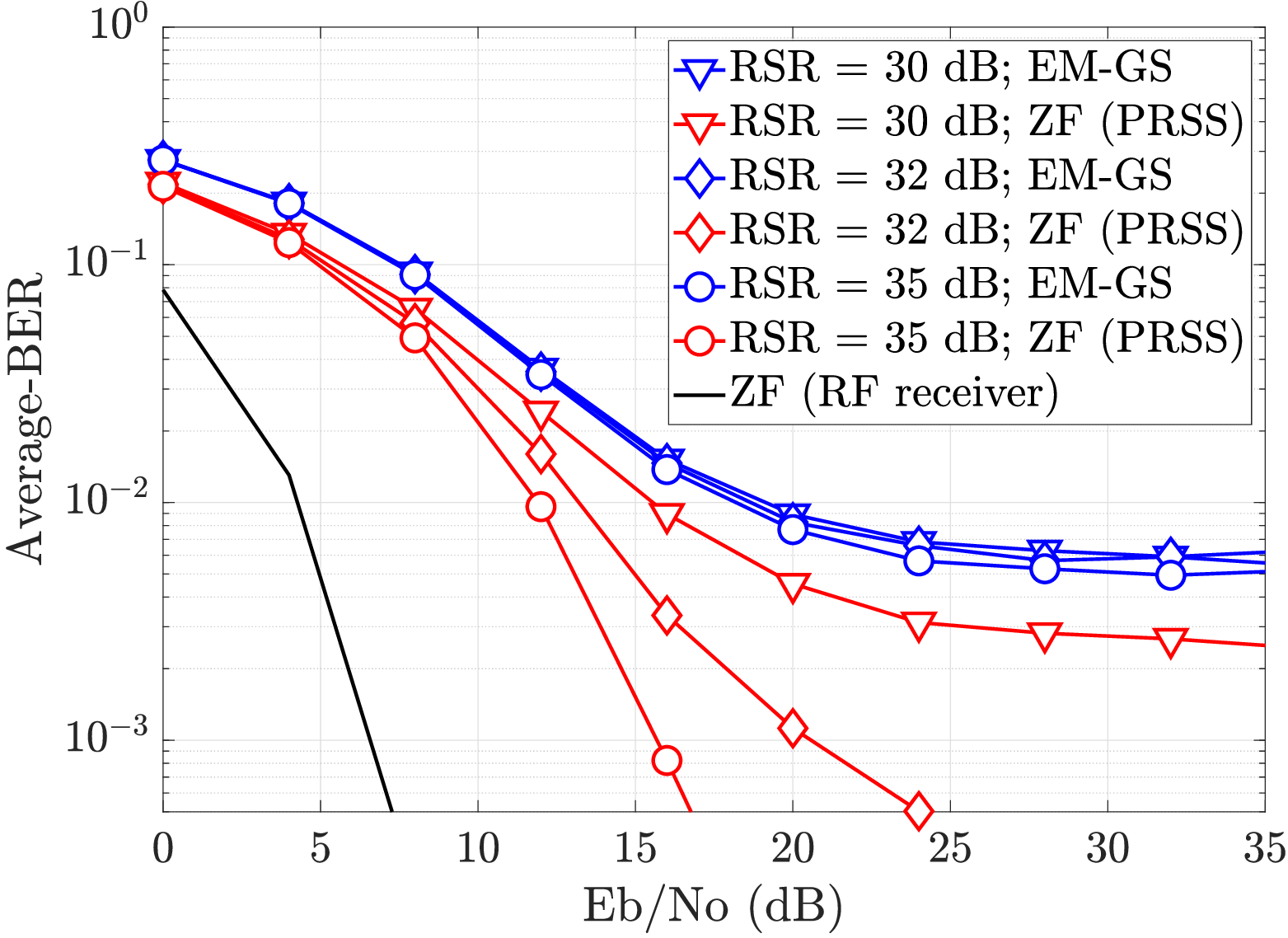}
	\caption{\label{fig04} Comparison between PRSS and benchmarks in large MIMO systems: $M = 128$, $N = 64$. ZF (PRSS) consistently outperforms EM-GS across all achievable RSR values, with performance improving at higher RSR levels.}
	\vspace{-1em}
\end{figure}
This study examines the detection performance of PRSS in large-scale MIMO configurations: $M = 128$, $N = 64$.
ML detection is not presented due to prohibitive computational complexity.
\figref{fig04} demonstrates that ZF (PRSS) detection exhibits sensitivity to RSR values, with higher RSR yielding superior performance.
This behavior aligns with the results in in \thmref{thm01} and \textit{Case Study 1}.
Across all examined RSR values, ZF (PRSS) consistently delivers superior detection performance compared to EM-GS.
When compared to conventional systems, ZF (RF receiver) shows approximately $10$ dB performance advantage over ZF (PRSS).
However, RA receivers provide more than $20$ dB sensitivity gain compared to RF receivers \cite{Bussey2022}.
Furthermore, higher RSR values become necessary as the number of users increases.

\section{Conclusion and Outlook}
This letter presents PRSS, a transmitter-receiver co-design solution that addresses RA-MIMO scalability challenges by transmitting each symbol with optimal $\pi/2$ phase rotation across two time slots.
This approach transforms the nonlinear RA-MIMO problem into an equivalent linear RF-MIMO model, enabling direct application of conventional detection algorithms.
Simulation results demonstrate substantial performance gains: $2.5$ dB improvement with ML detection and $10$ dB with ZF detection over existing single-transmission methods.
Additionally, ZF complexity is one order of magnitude lower than EM-GS, marking substantial progress toward practical scalable RA-MIMO systems.

Several research directions remain for future work. 
First, investigating the optimal phase offsets under more comprehensive noise sources, including blackbody radiation noise and quantum shot noise, warrants further exploration.
Second, extending PRSS to RA-MIMO channel estimation and downlink precoding.
Third, studying practical considerations including channel time-variance, dual-slot latency, and synchronization sensitivity.
Fourth, developing techniques to reduce the high RSR requirements of PRSS for broader applicability.

\appendices
\section{Proof of \thmref{thm01}} \label{appthm01}
The expressions of $y_{m}^{(1)}$ and $y_{m}^{(2)}$ in \eqref{eqn10190401} can be reformulated as the following linear system
\begin{equation} \label{eqn10270327}
	\begin{bmatrix}
		y_{m}^{(1)} \\
		y_{m}^{(2)}
	\end{bmatrix}
	=
	\mathbf{A}_{m}(\phi)
	\begin{bmatrix}
		\Re\{s_{m}\} \\
		\Im\{s_{m}\}
	\end{bmatrix}
	+
	\begin{bmatrix}
		\Re\{u_{m}v_{m}^{(1)}\}\\
		\Re\{u_{m}v_{m}^{(2)}\}
	\end{bmatrix},
\end{equation}
where $\mathbf{A}_{m}(\phi)$ is the measurement matrix, as follows
\begin{equation} \label{eqn08100327}
	\mathbf{A}_{m}(\phi) = 	
	\begin{bmatrix}
		\Re\{u_{m}\} & -\Im\{u_{m}\} \\
		\Re\{u_{m}e^{j\phi}\} & -\Im\{u_{m}e^{j\phi}\}
	\end{bmatrix}.
\end{equation}
The LS solution of \eqref{eqn10270327} for any invertable $\mathbf{A}_{m}(\phi)$ is as follows 
\begin{equation} \label{eqn08290327}
	\begin{bmatrix}
		\Re\{\hat{s}_{m}\} \\
		\Im\{\hat{s}_{m}\}
	\end{bmatrix}
	=\big[\mathbf{A}_{m}(\phi)\big]^{-1}
	\begin{bmatrix}
		y_{m}^{(1)} \\
		y_{m}^{(2)}
	\end{bmatrix}.
\end{equation}
For any noise distribution with zero mean and variance $\sigma_{v}^{2}$, the LS estimator has the following mean square error
\begin{equation} \label{eqn08390619}
	E[|\hat{s}_m - s_m|^2] = \sigma_v^2 \mathrm{Tr}([\mathbf{A}_m(\phi)^T \mathbf{A}_m(\phi)]^{-1}),
\end{equation}
where $\mathrm{Tr}(\cdot)$  denotes the trace operator.
To find the optimal phase $\phi$, we compute the Gram matrix from \eqref{eqn08100327}
\begin{equation}
	\mathbf{A}_m(\phi)^T \mathbf{A}_m(\phi) = |u_m|^2 \begin{bmatrix} 1 & \cos(\phi) \\ \cos(\phi) & 1 \end{bmatrix}.
\end{equation}
The trace of the inverse Gram matrix evaluates to
\begin{equation}
	\mathrm{Tr}([\mathbf{A}_m(\phi)^T \mathbf{A}_m(\phi)]^{-1}) = \frac{2}{|u_m|^2 \sin^2(\phi)}.
\end{equation}
To minimize the LS estimation error, we must minimize this trace expression.
Since the denominator contains $\sin^2(\phi)$, the minimum occurs when $|\sin(\phi)|$ is maximized, i.e., when $|\sin(\phi)| = 1$.
This condition is satisfied at $\phi^{\star} = \pm\frac{\pi}{2}$.
For the case $\phi^{\star} = \frac{\pi}{2}$, substituting the $\mathbf{A}(\frac{\pi}{2})$ into \eqref{eqn08290327} and converting to complex form via $\hat{s}_{m} = \Re\{\hat{s}_{m}\} + j \Im\{\hat{s}_{m}\}$ yields \eqref{eqn12060304a} in \thmref{thm01}.
The derivation for $\phi^{\star} = -\frac{\pi}{2}$ follows the same procedure and yields equivalent performance.

\bibliographystyle{myIEEEtran.bst}
\bibliography{../IEEEabrv,../thesis_list}

\ifCLASSOPTIONcaptionsoff
\newpage
\fi

\end{document}

%% file: Acronyms.tex
\newacronym{1d}{1D}{one-dimensional}
\newacronym{2d}{2D}{two-dimensional }
\newacronym{3d}{3D}{three-dimensional}
\newacronym{3gpp}{3GPP}{3rd Generation Partnership Project}
\newacronym{5g}{5G}{fifth-generation}
\newacronym{6g}{6G}{sixth-generation}

\newacronym{ap}{AP}{access point}
\newacronym{amp}{AMP}{approximate message passing}
\newacronym{ask}{ASK}{amplitude shift keying}
\newacronym{awgn}{AWGN}{additive white Gaussian noise}

\newacronym{ber}{BER}{bit error rate}
\newacronym{bfgs}{BFGS}{Broyden–Fletcher–Goldfarb–Shanno}
\newacronym{bler}{BLER}{bit error rate}
\newacronym{bp}{BP}{belief propagation}
\newacronym{bpsk}{BPSK}{binary phase shift keying}
\newacronym{bs}{BS}{base station}

\newacronym{cbsm}{CBSM}{correlation-based stochastic model}
\newacronym{csirs}{CSI-RS}{channel state information reference signal}
\newacronym{cdf}{CDF}{cumulative distribution function}
\newacronym{cdma}{CDMA}{code division multiple access}
\newacronym{cg}{CG}{conjugate gradient descent}
\newacronym{clt}{CLT}{central limit theorem}
\newacronym{csi}{CSI}{channel state information }
\newacronym{cvp}{CVP}{closest vector problem}

\newacronym{dof}{DoF}{degrees of freedom}
\newacronym{dsqt}{DSQT}{dual-slot quadrature transmission}

\newacronym{edw}{EDW}{exponentially decaying window}
\newacronym{eit}{EIT}{electromagnetically induced transparency}
\newacronym{elaa}{ELAA}{extremely large aperture array}
\newacronym{etsi}{ETSI}{European Telecommunications Standards Institute}

\newacronym{ff}{FF}{far-field}
\newacronym{fcsd}{FCSD}{fixed-complexity sphere decoder}
\newacronym{fec}{FEC}{forward error correction}
\newacronym{fspl}{FSPL}{free space path loss}

\newacronym{gbsm}{GBSM}{geometry-based stochastic model}
\newacronym{gd}{GD}{gradient descent}
\newacronym{gmsk}{GMSK}{Gaussian minimum shift keying}
\newacronym{gs}{GS}{Gauss-Seidel}
\newacronym{gsm}{GSM}{global system for mobile communication}

\newacronym{iid}{i.i.d.}{independently and identical distributed}
\newacronym{ils}{ILS}{integer least-squares}
\newacronym{ind}{i.n.d.}{independently and non-identical distributed}
\newacronym{imt}{IMT}{International Mobile Telecommunications}
\newacronym{isac}{ISAC}{integrated sensing and communication}
\newacronym{isi}{ISI}{intersymbol interference}
\newacronym{itur}{ITU-R}{International Telecommunication Union Radiocommunication Sector}
\newacronym{iui}{IUI}{inter-user interference}

\newacronym{ji}{JI}{Jacobi iteration}
\newacronym{jsac}{JSAC}{joint sensing and communication}

\newacronym{las}{LAS}{likelihood ascent search}
\newacronym{lbfgs}{LBFGS}{limited-memory Broyden–Fletcher–Goldfarb–Shanno}
\newacronym{leo}{LEO}{low Earth orbit}
\newacronym{lll}{LLL}{Lenstra-Lenstra-Lov\'{a}sz}
\newacronym{llr}{LLR}{log-likelihood ratio}
\newacronym{los}{LoS}{line of sight}
\newacronym{lo}{LO}{local oscillator}
\newacronym{lr}{LR}{lattice reduction}
\newacronym{lmmse}{LMMSE}{minimum mean square error}
\newacronym{ls}{LS}{least square}
\newacronym{lsd}{LSD}{list sphere decoder}
\newacronym{lte}{LTE}{long-term evolution}

\newacronym{map}{MAP}{maximum a posteriori}
\newacronym{mf}{MF}{matched filter}
\newacronym{mimo}{MIMO}{multiple-input multiple-output}
\newacronym{mld}{MLD}{maximum likelihood detection}
\newacronym{ml}{ML}{maximum likelihood}
\newacronym{mmimo}{mMIMO}{massive multiple-input multiple-output}
\newacronym{mrc}{MRC}{maximum ratio combining}
\newacronym{mse}{MSE}{mean square error}
\newacronym{ms}{MS}{matrix-splitting}
\newacronym{mt}{MT}{mobile terminal}

\newacronym{nf}{NF}{near-field}
\newacronym{nlos}{NLoS}{non-LoS}

\newacronym{od}{OD}{orthogonality defect}

\newacronym{pdf}{PDF}{probability distribution function}
\newacronym{pda}{PDA}{probabilistic data association}
\newacronym{pep}{PEP}{pairwise error probability}
\newacronym{pl}{PL}{path-loss}
\newacronym{prss}{PRSS}{phase-rotated symbol spreading}
\newacronym{pr}{PR}{phase retrieval}
\newacronym{pmf}{PMF}{probability mass function}
\newacronym{pwm}{PWM}{plane-wave model}

\newacronym{qam}{QAM}{quadrature amplitude modulation}
\newacronym{qpsk}{QPSK}{quadrature phase shift keying}
\newacronym{qn}{QN}{quasi-Newton}

\newacronym{rts}{RTS}{reactive tabu search}
\newacronym{ri}{RI}{Richardson iteration}
\newacronym{ris}{RIS}{reconfigurable intelligent surface}
\newacronym{rf}{RF}{radio-frequency}
\newacronym{rpe}{RPE}{rotational phase encoding}
\newacronym{rsr}{RSR}{reference-to-signal ratio}
\newacronym{rss}{RSS}{received signal strength}
\newacronym{rzf}{RZF}{regularized-ZF}

\newacronym{sa}{SA}{Seysen's algorithm}
\newacronym{sd}{SD}{steepest descent}
\newacronym{sdr}{SDR}{semidefinite relaxation}
\newacronym{ser}{SER}{symbol error rate}
\newacronym{sf}{SF}{shadow fading}
\newacronym{sic}{SIC}{succesive interference cancellation}
\newacronym{sinr}{SINR}{signal-to-interference-plus-noise ratio}
\newacronym{siso}{SISO}{single input single output}
\newacronym{snr}{SNR}{signal-to-noise ratio}
\newacronym{sns}{SNS}{spatial non-stationarity}
\newacronym{sota}{SoTA}{state-of-the-art}
\newacronym{ssor}{SSOR}{symmetric successive over-relaxation}
\newacronym{svd}{SVD}{singular value decomposition }
\newacronym{swm}{SWM}{spherical-wave model}

\newacronym{tr}{TR}{Technical Report}
\newacronym{ts}{TS}{tabu search}

\newacronym{uca}{UCA}{uniform cylindrical array}
\newacronym{ue}{UE}{user equipment}
\newacronym{ula}{ULA}{uniform linear array}
\newacronym{ura}{URA}{uniform rectangular array}
\newacronym{uma}{UMa}{urban macro}
\newacronym{umi}{UMi}{urban micro}
\newacronym{upa}{UPA}{uniform planar array}
\newacronym{ut}{UT}{user terminal}

\newacronym{vblast}{V-BLAST}{vertical Bell Labs layered space-time}

\newacronym{xlmimo}{XL-MIMO}{extra-large multiple-input multiple-output}

\newacronym{zf}{ZF}{zero-forcing}